\documentclass[12pt]{iopart}

\usepackage{iopams}  
\usepackage{amssymb}
\usepackage{graphicx}
\begin{document}

\title[Event-by-event mean $p_{\rm T}$ fluctuations and transverse size of color flux tube]
{Event-by-event mean $p_{\rm T}$ fluctuations and transverse size of color flux tube generated 
in $p$-$p$ collisions at $\sqrt{s}$=0.90TeV
}

\author{Takeshi Osada}
\address{Department of Physics, Faculty of Liberal Arts and Sciences, \\ 
Tokyo City University, Tamazutsumi 1-28-1, Setagaya-ku, Tokyo 158-8557, Japan}
\ead{osada@ph.ns.tcu.ac.jp}
\author{Masamichi Ishihara}
\address{Department of Human Life Studies, Koriyama Women's University, Koriyama, Fukushima 963-8503, Japan}
\vspace{10pt}
\begin{indented}
\item[]\today
\end{indented}

\begin{abstract}
We propose a novel phenomenological model of mean transverse momentum fluctuations based on the Geometrical Scaling hypothesis. 
Bose-Einstein correlations between two gluons generated from an identical color flux tube are taken into account as a source of the fluctuation. 
We calculate an event-by-event fluctuation measure $\sqrt{C_m}/\langle p_{\rm T}\rangle$ and show that ALICE data observed at $\sqrt s=$0.90 TeV for $p$+$p$ collisions are reproduced. 
By fitting our model to the experimental data, 
we evaluate the transverse size of the color flux tube as a function of the multiplicity. 
\end{abstract}
\section{Introduction}
Event-by-event fluctuation \cite{Voloshin:1999yf} (or correlation) of transverse momentum $p_{\rm T}$ 
is considered to be a prove of the hot and dense matter 
generated in high-energy $p$-$p$ and heavy-ion collisions theoretically \cite{Korus:2001fv,Gavin:2003cb,Gavin:2011gr} 
and have been extensively studied experimentally \cite{Adams:2005ka,Adamova:2008sx}.
In particular, we expect that fluctuations of dynamical origin are useful for screening of models for the evolution of the matter 
and understanding of the mechanism in multi-particle interaction \cite{Voloshin:2003ud,Ferreiro:2003dw,Broniowski:2005ae}. 
(Although not dealt with in this paper, fluctuations originating in the structure of protons \cite{Mantysaari:2017xcx,Mantysaari:2016jaz,Shen:2016zpp} are also important.) 
If mean transverse momentum $\langle p_{\rm T}\rangle$ 
fluctuations are purely statistical in event-by-event,  
it should decrease according $[ dn_{\rm ch}/dy ]^{-0.5}$, where $dn_{\rm ch}/dy$ 
is the charged particle density. 
Such $[ dn_{\rm ch}/dy ]^{-0.5}$ dependence on the $\langle p_{\rm T}\rangle$ fluctuation 
can be explained by a so-called dilution scenario\cite{Stefan:2011es} in which fluctuations are caused by superposition of independent particle production.
In contrast,  
if there are fluctuations of dynamical origin in the initial stage of the multi-particle interactions and if those traces remain in some observables, 
they must give some important information on the dynamics of quark and gluon mater at the initial stage of the reaction \cite{Trainor:2011jd,Gavin:2012if,Bozek:2017elk}. 

Recently, the ALICE Collaboration has published data on event-by-event 
fluctuation of $\langle p_{\rm T}\rangle$ of charged particles in $p+p$ and Pb+Pb collisions 
at Large Hadron Collider (LHC) energies \cite{Abelev:2014ckr, Heckel:2015swa,Stefan:2011es}.
We note that Bose-Einstein correlations (BEC) of gluons generated 
in the early stage of the hadronic collisions can be regarded as a possible origin of such fluctuations.  
In a grasma flux tube picture \cite{Dumitru:2008wn,Dusling:2009ar}, 
the color flux tubes stretch in the longitudinal direction  
and they decay into gluons with momenta almost 
in the transverse direction at the central rapidity region.  
In order to take into account the BEC of gluon pairs,  
we need to specify a gluon source function 
in the transverse plane. 
Here, we assume that transverse size of the 
color flux tube is given by a saturation momentum scale $Q_{\rm sat}$ 
appearing in Geometrical scaling (GS) hypothesis 
\cite{
Stasto:2000er,
McLerran:2014apa,
McLerran:2013una,
Praszalowicz:2013fsa, 
Praszalowicz:2015dta, 
Praszalowicz:2015hia}. 
The hypothesis is based on the saturation picture \cite{Kharzeev:2000ph,Kharzeev:2001gp,McLerran:2010uc} 
which bring a intuitive and clear understanding to multiple particle interaction. 
Especially, the GS hypothesis provide a good description of multiplicity and collision energy dependence on $\langle p_{\rm T}\rangle$
observed in hadronic and nuclear collisions \cite{McLerran:2014apa}. 
Then, it is meaningful to investigate whether such fluctuations 
are explained by the GS hypothesis with introducing some correlations between gluons emitted from 
glasma flux tube . 

This paper has two purposes. 
Firstly, it is to investigate whether the GS 
gives a good description of the experimental data 
not only on the multiplicity dependence of the mean $p_{\rm T}$ but also 
on the event-by-event $\langle p_T\rangle$ fluctuation. 
Secondly, it is to investigate what information can be drawn from the experimental data on the event-by-event fluctuations.  
This paper is organized as follows. 
In Section \ref{sec:2} we briefly review the GS hypothesis 
and introduce a Tsallis-type function as a specific form of the universal function ${\cal F}(\tau)$ used in our formulation. 
The multiplicity dependences of $Q_{\rm sat}$ and an effective interaction area $S_{\rm T}$, which plays a central role 
in our model, are determined to reproduce experimental data on $\langle p_{\rm T} \rangle$ 
as a function of the charged particle multiplicity observed 
at $\sqrt{s}=0.90$ TeV for $p$-$p$ collision \cite{Aamodt:2010my,Abelev:2013bla}. 
We determine the value of the parameter $q$ appearing in the Tsallis-type 
universal function ${\cal F}$ to reproduce data on single inclusive $p_{\rm T}$ spectra. 
Assuming that there exists BEC 
between two gluons generated in the same color flux tube, 
we introduce a possible gluon correlation function in Section \ref{sec:3}. 
Then, ALICE data on event-by-event fluctuation of $\langle p_T \rangle$ are compared by our model 
which is characterized by the saturation momentum $Q_{\rm sat}$ with assumptions:   
i)~local parton-hadron duality is applied and gluon correlations in the transverse momentum space survives until the final state of hadrons,     
ii)~the universal function of GS is a Tsallis-type power-law function, 
iii)~two gluon correlation function is a BEC-type function given as eq.(\ref{eq:BEC_correlation}) with eq.(\ref{eq:chaoticity}).  
We close with Section \ref{sec:4} containing the summary and some further discussion. 

\section{Saturation momentum and single inclusive spectra}\label{sec:2} 
The GS hypothesis states that differential distributions of charged particles produced in 
hadronic collisions depend only on a specific combination of $p_T$ and $Q_s$ 
called saturation scale. 
The GS requires the gluon differential distribution function is expressed 
by the specific combination referred as a scaling variable, 
\begin{eqnarray}
 \frac{1}{2\pi p_{\rm T}} \frac{d^2n_g}{dp_{\rm T}dy}= S_{\rm T} 
  {\cal F}\left(\tau \right), \label{eq:gluon_spectra}
\end{eqnarray}
where $\tau\equiv p_{\rm T}^2/Q_s^2$. 
Here, $S_{\rm T}$ is an overlap transverse area of the colliding hadrons 
(or it may regard as an effective interaction area) and ${\cal F}(\tau)$ 
is an universal function of $\tau$. 
The saturation scale $Q_s$ is an external intermediate energy scale \cite{Praszalowicz:2013fsa}
which is larger than a typical non-perturbative energy scale ($\sim 0.2$ GeV) 
and smaller than a perturbative energy scale ($\sim 10$ GeV). 
The scale $Q_s$ is given by a function of Bjorken $x$ as the following:  
\begin{eqnarray}
   Q_s^2(x) = Q_0^2 \left(\frac{x}{x_0}\right)^{-\lambda}, 
   \label{eq:saturation_momentum}
\end{eqnarray} 
where $x\equiv p_T/W$, $W$ is the c.m.s.~scattering energy $\sqrt{s}$,  
$Q_0$ is an arbitrary scale parameter of order 1 GeV/$c$ 
and $x_0$ is a constant of order $10^{-3}$. 
The validity of this GS hypothesis has been confirmed by abundant experimental results in 
various reactions \cite{Praszalowicz:2015dta, Praszalowicz:2015hia} and these facts 
support the correctness of the basic idea of saturation \cite{Kharzeev:2000ph}. 
The saturation picture can be intuitively explained by the 
saturation momentum $Q_{\rm sat}$ instead of the scale $Q_s$. 
Denoting $\tilde W \equiv x_0W$,  
$Q_{\rm sat}$ is defined as a solution of the following equation;  
\begin{eqnarray}
  Q_{\rm sat}^2 = Q_0^2 ~\left(\frac{Q_{\rm sat}}{\tilde W} \right)^{-\lambda}.
\end{eqnarray}
Here, $Q_{\rm sat}$ should be regarded as a typical transverse momentum 
for given $\tilde W$ (or given multiplicity class as seen later) \cite{McLerran:2013una}.  
Then, its $\tilde W$ dependence is given by    
\begin{eqnarray}
    Q_{\rm sat}(\tilde W) =  Q_0 \left(\frac{\tilde W}{Q_0}\right)^{\frac{\lambda}{2+\lambda}}, 
    \label{eq:Qs_W}   
\end{eqnarray} 
and the scaling variable $\tau$ is replaced by the following: 
\begin{eqnarray}
  \tau = \left(\frac{p_T}{Q_{\rm sat}}\right)^{2+\lambda}.
  \label{eq:tau}   
\end{eqnarray}
Gluons with characteristic transverse momentum $Q_{\rm sat}$ 
occupy an area roughly given by $1/Q_{\rm sat}^2$ 
and its scattering cross section 
probed by the momentum transfer $Q_{\rm sat}$ is $\sim \alpha_s (Q_{\rm sat})/Q_{\rm sat}^2$ 
\cite{Kharzeev:2001gp,McLerran:2010uc}. 
In this case, the gluon density is saturated at the density of the following eq.(\ref{eq:saturation}) 
\cite{McLerran:2014apa,McLerran:2013una,Praszalowicz:2013fsa, Praszalowicz:2015dta, Praszalowicz:2015hia,Kharzeev:2000ph,Kharzeev:2001gp,McLerran:2010uc}, 
a critical density at which gluon begin to overlap each other. 
It is obtained by dividing the cross sectional areas of proton $S_{\rm T}$ 
by the scattering cross section, that is,  
\begin{eqnarray}
   \frac{dn_{g}}{dy} \sim \frac{1}{\alpha_s (Q_{\rm sat})}~S_{\rm T}Q_{\rm sat}^2. \label{eq:saturation}
\end{eqnarray}

\begin{figure}
\centerline{\includegraphics[width=9.0cm]{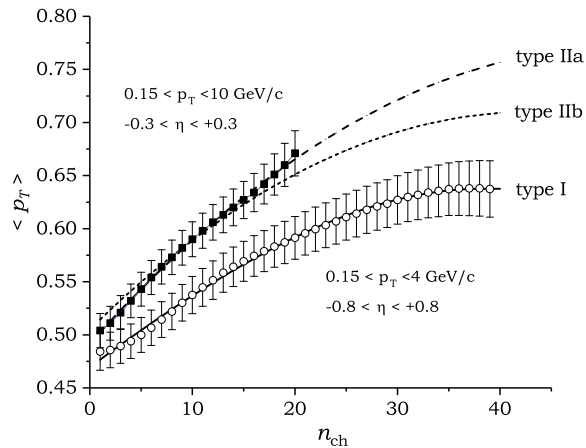}}
\caption{
The ALICE data on $\langle p_T \rangle$ as function of $n_{\rm ch}$ for $p$-$p$ 
collisions at energy $\sqrt{s}$ = 0.90 TeV.
The data displayed by open and closed circles are observed for two different interval of $p_T$ and $\eta$ of charged particles 
published in \cite{Aamodt:2010my} and \cite{Abelev:2013bla}, respectively.
The solid line is the (3.rd order polynomial) fitting the data of  \cite{Aamodt:2010my}, which is named type~I.  
It is extrapolated by an assumption that $\langle p_{\rm T}\rangle$ is constant for $n_{\rm ch}>40$. 
The long and short dashed lines are (2.nd oder polynomial) fittings the data of \cite{Abelev:2013bla} 
with two different extrapolations for $n_{\rm ch}>20$, named type IIa and IIb, respctively.} 
\label{fig:avpt_vs_nch_pp0.90}
\end{figure}

\begin{figure}
\centerline{\includegraphics[width=10.0cm]{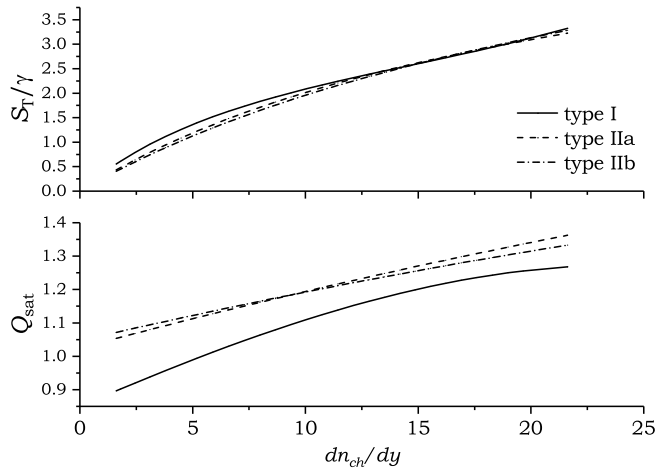}}
\caption{The multiplicity dependence of the $S_{\rm T}/\gamma$ and 
the saturation momentum $Q_{\rm sat}$ obtained for $p$-$p$ collisions at energy 
$\sqrt{s}$ = 0.90 TeV for each average interval type I and type II.}\label{fig:QsRt}
\end{figure}

\begin{figure}
\centerline{\includegraphics[width=10.0cm]{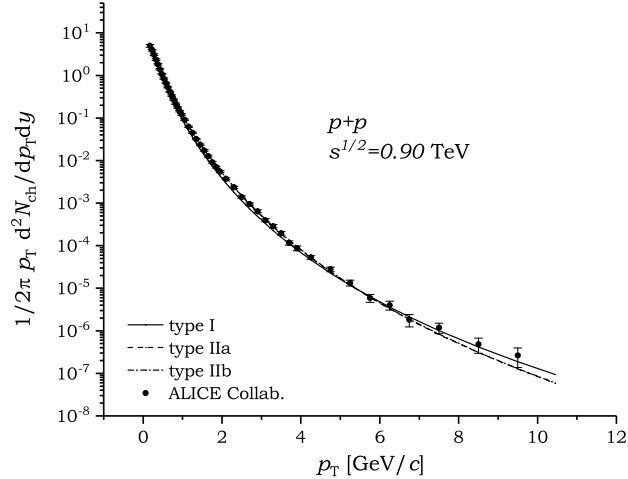}}
\caption{The $p_{\rm T}$ spectra fitted to 
the ALICE data for $p$-$p$ collisions at energy $\sqrt{s}$ = 0.90 TeV \cite{Aamodt:2010my}. 
The values of $q$ and $\kappa$, for each acceptance and extrapolation type I, IIa and IIb, are shown in Table \ref{tab:1}.
}\label{fig:single_pt_pp090}
\end{figure}

Since transverse momentum spectra 
observed in various hadronic collisions are well reproduced 
by so-called Tsallis-type distributions,  
we assume the form of the universal function appearing in the GS is 
the following Tsallis-type distribution \cite{McLerran:2014apa}: 
\begin{eqnarray}
{\cal F}(\tau) =\left[~ 1+(q-1) \frac{\tau^{1/(\lambda+2)}}{\kappa} ~\right]^{\frac{-1}{q-1}}, 
\label{eq:universal_func}
\end{eqnarray}
where $q$ and $\kappa$ are constants\footnote{
Note that using eq.(\ref{eq:universal_func}) with eq.(\ref{eq:tau}) leads 
a transverse spectrum proportional to 
$\exp(-p_{T}/T)$ in the limit of $q\to1$. 
}. 
Here, we assume local parton-hadron duality \cite{Azimov:1984np}, i.e., 
the gluon differential momentum distribution is proportional to that of hadrons: 
its proportional constant 
$\gamma$ satisfies 
$n_g/\gamma=n_{\rm ch}$. 
Then, the gluon's and charged hadron's transverse spectra are 
characterized by saturation momentum $Q_{\rm sat}$ as follows: 
\begin{eqnarray}
    \frac{1}{S_{\rm T}} 
    \frac{1}{2\pi p_{\rm T}} \frac{d^2n_g}{dp_{\rm T} dy} = 
    \frac{\gamma}{S_{\rm T}} 
    \frac{1}{2\pi p_{\rm T}} \frac{d^2n_{\rm ch}}{dp_{\rm T}dy} 
    = \left[1+(q-1)\frac{p_T}{\kappa Q_{\rm sat}(\tilde W)} \right]^{-1/(q-1)}. 
    \label{eq:Tsallis-type}
\end{eqnarray}
The $m$-th moment of the 
spectrum for a limiting transverse momentum range $[p_{T_{\rm mim}},p_{T_{\rm max}}]$ is proportional to 
the following integral: 
\begin{eqnarray}
   {\cal I}_{m} [p_{T_{\rm mim}},p_{T_{\rm max}}] &\equiv& 
   \int^{p_{T_{\rm max}}}_{p_{T_{\rm min}}} \!\! dp_T~ 
   p_T^{m} \left[1+(q-1)\frac{p_T}{\kappa Q_{\rm sat}} \right]^{\frac{-1}{q-1}} \nonumber \\
  &=& 
 \sum_{k=1}^{m+1} \frac{m! ~\prod_{i=1}^{k} {\cal G}^i}{(m+1-k)!}
 ~\bigg[ {\cal F}^{k}_{m+1-k} \bigg]^{p_{T_{\rm mim}}}_{p_{T_{\rm max}}},
 \label{eq:general_integral}
\end{eqnarray}
where
\begin{eqnarray}
  {\cal F}^{k}_{m}  &\equiv&
   p_T^{m}   
   \left\{1+(q-1)\frac{p_T}{\kappa Q_{\rm sat}} 
   \right\}^{\frac{-1}{q-1}+k}, 
   \quad {\cal G}^{k} \equiv \left[\frac{\kappa Q_{\rm sat}}{(k+1)-kq} \right].    
\end{eqnarray}
Thus, the charged hadron multiplicity density 
and mean transverse momentum, 
respectively, are given by \cite{McLerran:2010ex,Praszalowicz:2014kaa} 
\begin{eqnarray}
 \frac{dn_{\rm ch}}{dy} &=& 
 \frac{2\pi S_{\rm T}}{\gamma} {\cal I}_1 [p_{T_{\rm mim}},p_{T_{\rm max}}], \label{eq:full_expression_pt}\\ 
    \langle p_{\rm T} \rangle &=& {\cal I}_2/ {\cal I}_1, 
 \label{eq:full_expression_dndy} 
\end{eqnarray}
where the rapidity density is assumed to be constant in the central rapidity region. 
In the limit of  $p_{T_{\rm min}}\to0$ and $p_{T_{\rm max}}\to\infty$ 
($q<1+\frac{1}{m+1}$ is required to converge the integration in this case), 
we have
\begin{eqnarray}
   {\cal I}^{}_{m} [0,\infty] =
   m! \prod_{k=1}^{m+1}{\cal G}^{k} =   \frac{
   m!}{\prod_{k=1}^{m+1} \left[(k+1)-kq\right]} ~\left[\kappa Q_{\rm sat}\right]^{m+1}. 
   \label{eq:full_phase_space}
\end{eqnarray}
By using eq.(\ref{eq:full_phase_space}), 
eqs. (\ref{eq:full_expression_pt}) and (\ref{eq:full_expression_dndy}) 
can be reduced to the following explicit expressions 
\begin{eqnarray}
     \frac{dn_{\rm ch}}{dy} 
   =2\pi\frac{S_{\rm T}}{\gamma}\frac{[\kappa Q_{\rm sat}]^{2} }{ (2-q)(3-2q)},  \qquad 
   \langle p_T \rangle =
   \frac{ 2\kappa Q_{\rm sat} }{4-3q},  \label{eq:single_dndy_and_pt} 
\end{eqnarray} 
respectively. 
It may be worth to roughly estimate the values of quantities such as $Q_{\rm sat}$ by using eq.(\ref{eq:single_dndy_and_pt}) 
before detail analysis. 
(In accurate calculations as shown later in Fig.\ref{fig:QsRt},  
$Q_{\rm sat}$ and $S_{\rm T}$ are evaluated by eqs. (\ref{eq:full_expression_pt}) and (\ref{eq:full_expression_dndy}), 
where the finite interval  $[p_{T_{\rm mim}},p_{T_{\rm max}}]$ is taken into account for the evaluation of ~${\cal I}_m$.)
Let's evaluate $Q_{\rm sat}$ and $S_{\rm T}/\gamma$ for a typical event with multiplicity $n_{\rm ch} = 7.5$ 
which is nearly equarl to the value using in a single Negative Binomial Distribution fitting ($\langle n_{\rm ch}\rangle_{\rm NBD} = 7.49$)  
\cite{Ghosh:2012xh} for $p$+$p$ collisions at energy $\sqrt{s}=$0.90 TeV. 
The corresponding mean transverse momentum $\langle p_{\rm T} \rangle$ is approximately 0.51 GeV/$c$ \cite{Aamodt:2010my}.  
For $\kappa$=0.15 and $q\approx1+1/9\approx 1.1$ \cite{McLerran:2014apa}, the second equation of eq.(\ref{eq:single_dndy_and_pt}) leads to $Q_{\rm sat}=1.19$ GeV/$c$ 
and then the first equation of eq.(\ref{eq:single_dndy_and_pt}) leads to $S_{\rm T}/\gamma=1.05~{\rm fm}^2$. 
Note here that we obtain the saturation momentum $Q_{\rm sat}=1.19$ GeV/$c$ by tuning parameters in eq.(\ref{eq:Qs_W}): 
$Q_0=$1.0 GeV, $\lambda=0.22$ \cite{McLerran:2014apa} , $W=0.9$ TeV, and  $x_0=6.4\times10^{-3}$. 
The ALICE Collaboration presents data on $\langle p_{\rm T} \rangle$ as a function of $n_{\rm ch}$ for 
two different experimental window in $p_{T}$ and pseudo rapidity $\eta$ \cite{Aamodt:2010my,Abelev:2013bla}. 
One interval is $0.15~{\rm GeV/{\it c}} < p_{\rm T} <4.0~{\rm GeV/{\it c}}$, $|\eta|<0.8$~\cite{Aamodt:2010my}  (denoted by type I) 
and the other is $0.15~{\rm GeV/{\it c}} <p_{\rm T} <10~{\rm GeV/{\it c}}$, $|\eta|<0.3$ \cite{Abelev:2013bla} 
(denoted by type IIa and IIb, which are two possible extrapolations of different type).   
Both type I and type II data indicate that $\langle p_{\rm T} \rangle$ monotonically increases 
as $n_{\rm ch}$ increase. The experimental data 
are interpolated and extrapolated as shown by solid curves in Fig.\ref{fig:avpt_vs_nch_pp0.90} for use of the numerical evaluation of $Q_{\rm sat}$, 
$S_{\rm T}/\gamma$, {\it etc}. 
Those two quantities are simultaneously determined as a function of $n_{\rm ch}$, 
if values of $q$ and $\kappa$ are given respectively.  
Figure \ref{fig:QsRt} shows the $dn_{\rm ch}/dy$ dependence of the saturation momentum 
$Q_{\rm sat}$ and the effective interaction area $S_{\rm T}/\gamma$, 
where parameters $q$ and $\kappa$ are fixed to reproduce experimental data on 
transverse momentum spectra (as shown in Fig.\ref{fig:single_pt_pp090}). 
The single particle inclusive transverse momentum spectrum is obtained by superposition of 
$p_{\rm T}$ spectra over wide event multiplicity class. 
The multiplicity distribution of charged hadrons observed at  
$\sqrt{s}=$0.90 TeV \cite{Aamodt:2010ft,Adam:2015gka} can be reproduced by a 
negative binomial distribution \footnote{
It is interesting to note that the gluon multiplicity distribution is a negative binomial 
distribution if gluons are emitted from independent Glasma flux tubes \cite{Gelis:2009wh}.}
\begin{eqnarray}
  P_{\rm NBD}(n_{\rm ch},\langle n_{\rm ch} \rangle , k)= \frac{\Gamma(k+n_{\rm ch})}{\Gamma(k)\Gamma(n_{\rm ch}+1)} 
  \frac{\langle n_{\rm ch}\rangle^{n_{\rm ch}} k^k}{(\langle n_{\rm ch} \rangle+k)^{n_{\rm ch}+k}}. 
  \label{eq:NBD} 
\end{eqnarray} 
In order to accurately reproduce the multiplicity distribution, we use a weighted sum of two NBD functions 
employed by ALICE collaboration:  
\begin{eqnarray}
   P(n_{ch}) = 
   \lambda_{n} \Big[ \alpha  P_{\rm NBD} (n_{\rm ch}, \langle n_{\rm ch}\rangle_1 , k_1) 
   + (1-\alpha)  P_{\rm NBD}(n_{\rm ch}, \langle n_{\rm ch} \rangle_2 , k_2) \Big], 
   \label{eq:double_NBD}
\end{eqnarray} 
where double-NBD fit parameters used, such as $\langle n_{\rm ch}\rangle_1$ and $k_1$,   
are found in the Table 9 in Ref.\cite{Adam:2015gka}. 
Hence the charged particle transverse momentum spectra is given by 
\begin{figure}
\centerline{\includegraphics[width=10.0cm]{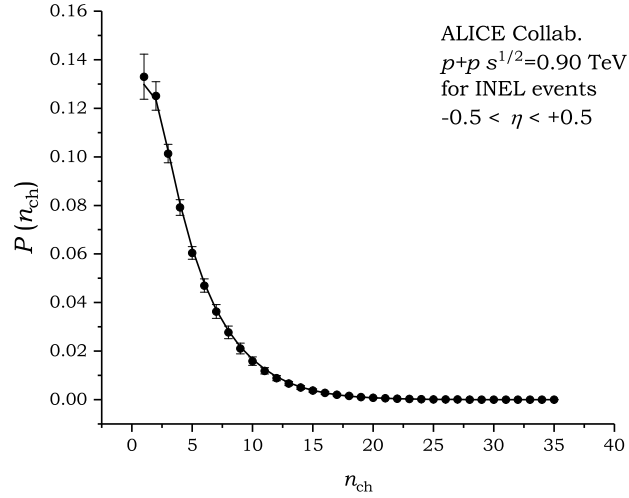}}
\caption{
The multiplicity distribution for $p$-$p$ collisions at 
energy $\sqrt{s}$ = 0.90 TeV observed by ALICE Collaboration \cite{Adam:2015gka}. 
The solid curve is a result of fit with eq.(\ref{eq:double_NBD}) to the data. 
}\label{fig:pn_nbd}
\end{figure}
\begin{eqnarray}
    \frac{1}{2\pi p_T}
    \frac{d^2N_{\rm ch}}{dp_Tdy} = 
    \sum_{n_{\rm ch}} P(n_{\rm ch}) ~\frac{S_{\rm T}(n_{\rm ch})}{\gamma}
    \left[1+(q-1)\frac{p_T}{\kappa Q_{\rm sat}(n_{\rm ch})} \right]^{-1/(q-1)},
    \label{eq:convolution_model}
\end{eqnarray} 
where $S_{\rm T}/\gamma$ is re-normalized 
to reproduce particle density $dN_{\rm ch}/d\eta=3.02$ for 
inelastic (INEL) $p+p$ collisions at 0.90 TeV \cite{Aamodt:2010ft}. 
The resultant transverse momentum spectra is shown in Fig.\ref{fig:single_pt_pp090}. 
As shown in Fig.\ref{fig:single_pt_pp090}, 
the transverse spectra is well-fitted by eq.(\ref{eq:convolution_model}) with $q= 1.106-1.124$, $\kappa=0.135-0.138$.  
In $p_{\rm T}>8.0$ GeV/c region, numerical results of the type IIa and IIb are slightly smaller than the experimental data. 
Since there are no substantial difference between the result of type IIa and IIb, 
this slight under estimation is not due to the difference in extrapolation in the region with large multiplicity ($n_{\rm ch}>20$). 
It is probably due to the multiplicity dependence of $\langle p_{\rm T} \rangle$ in the small multiplicity (large transverse momentum) region.

\begin{table}
\begin{center}
\caption{
Values of $q$ and $\kappa$ in eq.(\ref{eq:convolution_model}) obtained by fitting to the ALICE data\cite{Aamodt:2010my}.
}
\label{tab:1}
\begin{indented}
\item[] \begin{tabular}{ l c c c } 
\br 
~type~ & ~~$q$~~ & ~~$\kappa$~~ & $\chi2$/dof \\ \mr
I         & 1.124    & 0.1388    &  617/44 \\ 
IIa       & 1.106    & 0.1354    &  107/44 \\ 
IIb       & 1.106    & 0.1373    &  120/44 \\ 
\br 
\end{tabular}
\end{indented}
\end{center} 
\end{table}

\section{Event-by-event fluctuation of $p_{T}$ and size of color flux tube}\label{sec:3} 
\begin{figure}
\centerline{\includegraphics[width=12.0cm]{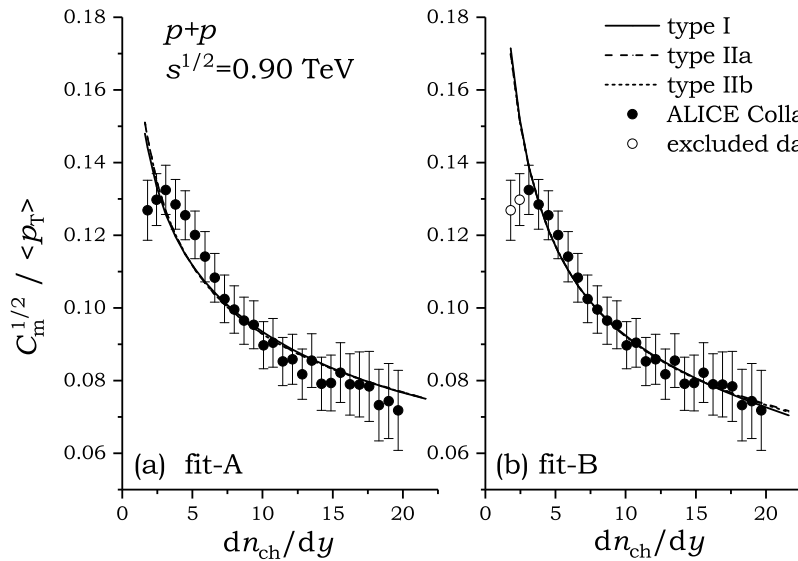}}
\caption{
Comparison of experimental data on $\sqrt{C_m}/\langle p_{\rm T}\rangle$ 
for $p$-$p$ collisions at energy $\sqrt{s}$ = 0.90 TeV \cite{Abelev:2014ckr} 
and model calculations based on eqs(\ref{eq:BEC_correlation}) and (\ref{eq:chaoticity}).  
It is assumed that $dn_{\rm ch}/d\eta \approx dn_{\rm ch}/dy$.  
a) Left panel: All 27 data points are used in fitting: 
b) Right panel: Two data points of $dn_{\rm ch}/dy<3.0$ are excluded in fitting.  
}\label{fig:CmPt2}
\end{figure}

The expression eq.(\ref{eq:convolution_model}) for the single inclusive $p_{\rm T}$ spectrum 
can be extended to the two particle inclusive distribution providing two particle correlation function.   
After high-energy hadron-hadron collisions, 
we consider color flux tubes that stretch between the receding hadrons. 
Their transverse size is expected to be order in $1/Q_{\rm sat}$ from the GS picture. 
Then, the subsequently multi-particle interaction proceeds by gluons generated by decay of the flux tubes \cite{Gavin:2008ev,Bzdak:2013zma}.  
If two gluons are generated from the identical flux tube, 
they are correlated with each other in the transverse momentum space \cite{Lappi:2010cp,Lappi:2009xa}.  
As a possible type of such correlation function, we use it commonly found in Bose-Einstein correlation analysis. 
When two identical bosons are generated from the tube with the transverse size scale $1/Q_{\rm sat}$, it may given by 
\begin{eqnarray}
  C({\bf p}_{T_1}, {\bf p}_{T_2}) \equiv 
  \frac{\displaystyle {\frac{d^4n_{\rm ch}}{d{\bf p}^2_{\rm T_1} d{\bf p}^2_{\rm T_2}}}}
  {\displaystyle \frac{d^2n_{\rm ch}}{d{\bf p}^2_{\rm T_1}}     \frac{d^2n_{\rm ch}}{d{\bf p}^2_{\rm T_2}} }
  =
 1+\lambda_g \exp \left(-\frac{({\bf p}_{\rm T_1}-{\bf p}_{\rm T_2})^2}{\sigma Q_{\rm sat}^2}\right),  \label{eq:BEC_correlation}
\end{eqnarray}
where $\lambda_g$ is an effective correlation strength, and $\sigma$ is introduced to accurately specify (or to modify) the correlation length. 
We here employ a Gaussian-type correlation function and ignore the longitudinal correlation \cite{He:2017laa} 
for the simplicity\footnote{
Even considering correlation of longitudinal momentum, its contribution is absorbed by the parameter $\lambda_g$ 
because we assume the Gaussian-type correlation function eq.(\ref{eq:BEC_correlation}).}.
The correlation strength $\lambda_g$ can be estimated by a ratio of a number of the identical gluon pairs 
generated from the same color flux tube to the total number of the identical gluon pairs without distinguishing 
from which tubes they were generated. 
Let $m_{\rm tube}$ is the average number of tubes and $k$ is the mean gluon multiplicity produced in 
each tube. Then, the total number of identical gluon (i.e., they have the same degree of freedom of color) 
pairs is given by ${}_{m_{\rm tube} k}{\rm C}_2$. 
On the other hand, the number of gluon pairs generated from the same tube is $m_{\rm tube}~{}_k {\rm C}_2$. 
Thus, we obtain the ratio as 
\begin{eqnarray}
   \lambda_g  \approx  \frac{k-1}{m_{\rm tube} k-1}.  
   \label{eq:lambda_model}
\end{eqnarray}   
Noting that the number of tubes can be estimated by  $m_{\rm tube} \sim S_{\rm T}/\frac{1}{\sigma Q_{\rm sat}^2}$,  
one may parametrize the correlation strength $\lambda_g$ as 
\begin{eqnarray}
\lambda_g =C~[\sigma S_{\rm T} Q_{\rm sat}^2]^n, \label{eq:chaoticity}
\end{eqnarray} 
where $n$ is an exponent parameter which is phenomenologically introduced and it is expected $n\approx-1$, if $k\gg1$. 
We also introduce a constant $C$ to eq.(\ref{eq:chaoticity}), assuming that contributions to the correlation strength of some other factors attribute to the constant. 
It is interesting to note that $S_{\rm T} Q_{\rm sat}^2\propto dn_{\rm ch}/dy$ and 
we then naturally reproduce the multiplicity dependence of the fluctuation of the mean transverse momentum 
\footnote{Recall the experimental data approximately behaves as $[dn_{\rm ch}/d\eta]^{-0.5}$, 
which is roughly explained by the dilution scenario in $p$-$p$ collisions. However, we also expect that 
$n\ne-0.5$ in Pb+Pb collisions \cite{Abelev:2014ckr}.}, i.e.,
$\sqrt{C_m}/\langle p_{\rm T}\rangle \propto \left[ dn_{\rm ch}/dy \right]^{n/2}$ and the dilution model is reproduced when $n=-1$. 
In the subsequent analysis, $C$ of equation (\ref{eq:chaoticity}) is set to $C$ = 1, for simplicity. 
If the gluon's correlation comes down to the hadrons observed after the freeze out, 
the residual correlation can be found in $p_{\rm T}$ fluctuation measure \cite{Abelev:2014ckr,Heckel:2015swa,Stefan:2011es}
\begin{eqnarray}
  C_m=\frac{\int\! d^2{\bf p}_{\rm T_1}\int\! d^2{\bf p}_{\rm T_2}}{m(m-1)}~
  \frac{d^4n_{\rm ch}}{d{\bf p}^2_{\rm T_1}  d{\bf p}^2_{\rm T_2}}~ 
  (p_{\rm T_1}-\langle p_{\rm T}\rangle)(p_{\rm T_2}-\langle p_{\rm T}\rangle), 
  \label{eq:Cm} 
\end{eqnarray}  
for event class with fixed multiplicity $m=\frac{dn_{\rm ch}}{d\eta} \times |\Delta\eta|$ 
and $ |\Delta\eta|$ is pseudo rapidity interval used in the observation. 
In Fig.\ref{fig:CmPt2}, we show two fit results for the fluctuation measure $\sqrt{C_m}/\langle p_{\rm T}\rangle$ observed by ALICE \cite{Aamodt:2010jj}. 
One is fit-A (the left panel), in which we use all 27 data points observed,  
and another is fit-B (the right panel), in which we exclude two data points of $dn_{\rm ch}/dy < 3.0$. 
Our simple model calculations nicely reproduce the experimental data except for small $dn_{\rm ch}/d\eta$ region.      
The best-fit parameter values of $n$ and $\sigma$ in eqs (\ref{eq:BEC_correlation}) and (\ref{eq:chaoticity}) are shown in Table \ref{tab:2}. 
As shown results in the Table \ref{tab:2}, the fit results suggest evidence in favor of  $n>-1$. 
Compared with the case of $n=-1$, decrease in $\lambda_g$ is suppressed in the case of $n>-1$ 
because the number of $m_{\rm tube}$ is substantially suppressed in the high multiplicity region. 
In our model, the power-low dependence of $dn_{\rm ch}/dy$ observed in $\sqrt{C_m}/\langle p_{\rm T}\rangle$ 
is derived from eq.(\ref{eq:lambda_model}). 

\begin{figure}
\centerline{\includegraphics[width=10.0cm]{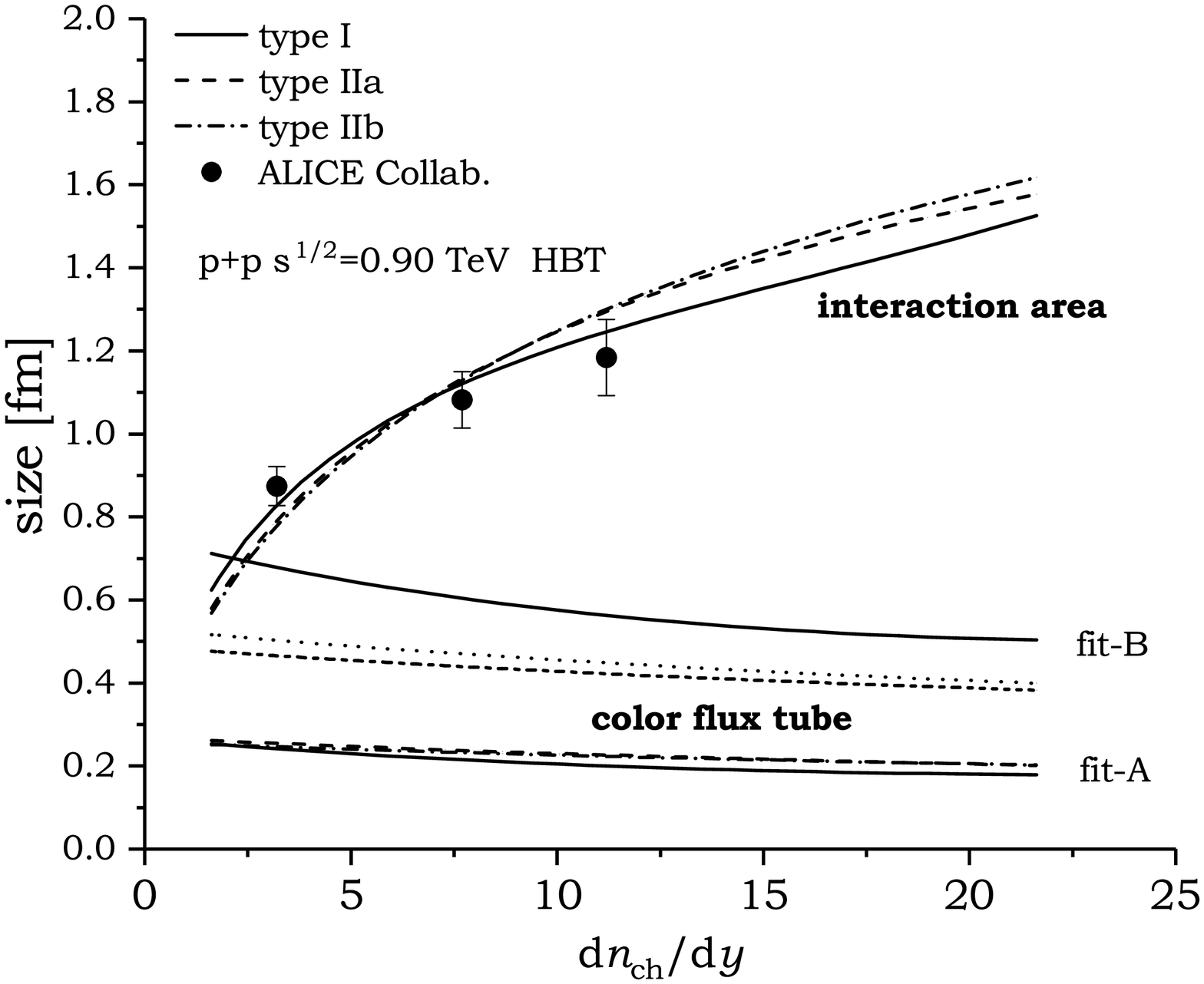}}
\caption{Effective radius of the interaction area $R_{\rm T}\equiv \sqrt{S_{\rm T}/\pi}$ and 
size of color flux tube $r_{\rm tube}\equiv 1/\sqrt{\sigma}Q_{\rm sat}$ for $p$-$p$ collisions at energy $\sqrt{s}$ = 0.90 TeV.  
We also show HBT radii \cite{Aamodt:2010jj} observed in the same energy 
to compare with our model results on $R_{\rm T}$. 
}\label{fig:tube_size}
\centerline{\includegraphics[width=10.0cm]{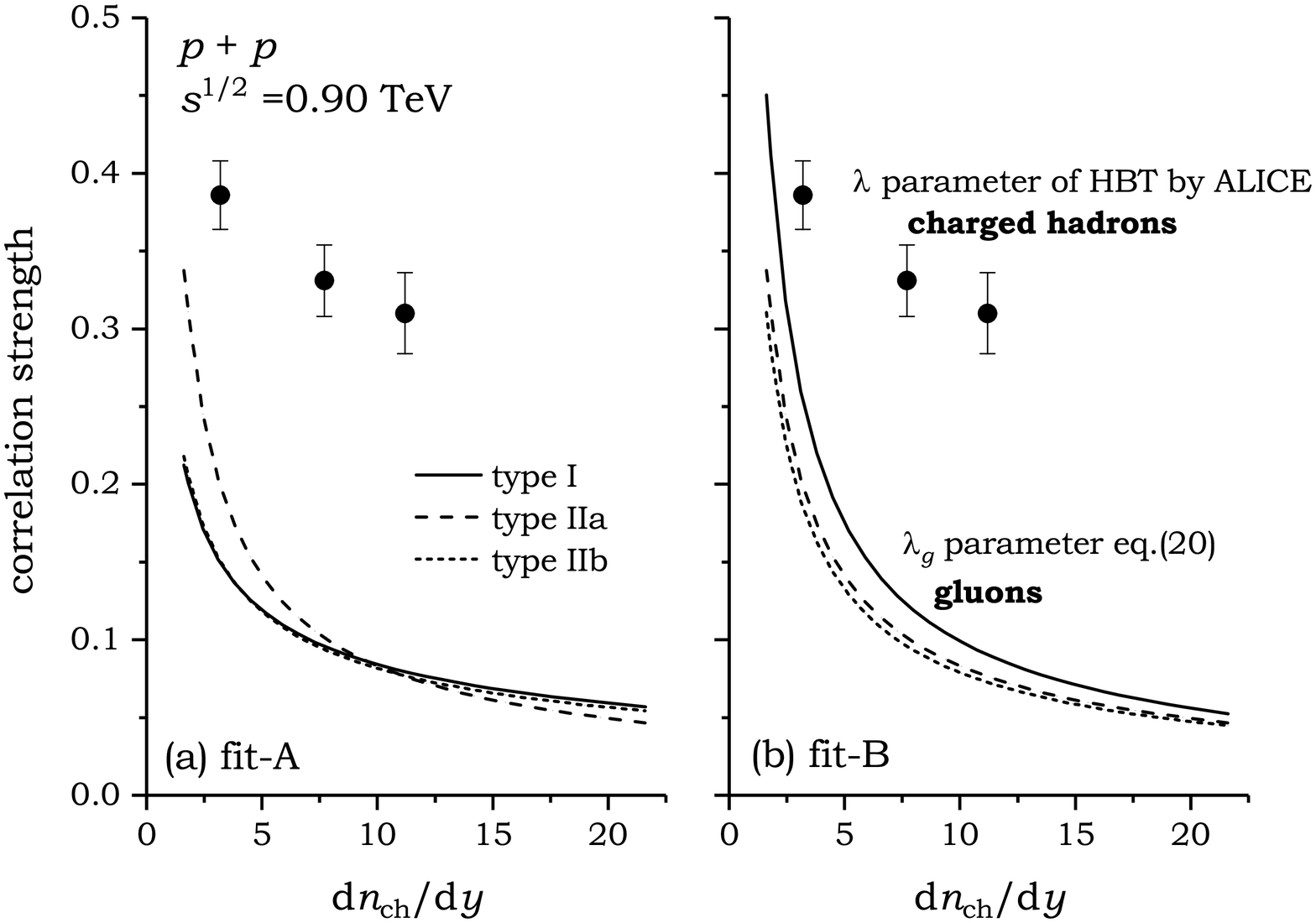}}
\caption{The correlation strength $\lambda_g$ obtained in our model for fit-A (in the left panel) and fit-B (in the right panel).  
We also show results of HBT measurements of the correlation strength $\lambda_{\rm HBT}$ parameter by ALICE for charged particles \cite{Aamodt:2010jj} for reference.}\label{fig:lambda} 
\end{figure}

The flux-tube size $r_{\rm tube}$ evaluated from $1/\sqrt{\sigma}Q_{\rm sat}$ and effective radius of the interaction area, which is defined by $R_{\rm T}\equiv\sqrt{S_{\rm T}/\pi}$, 
are shown in Fig.\ref{fig:tube_size}. 
As multiplicity increases, $R_{\rm T}$ also increases while $r_{\rm tube}$ decreases slightly. 
It is interesting to note that there is no substantial difference in the evaluation of $r_{\rm tube}$ between type I and type II (both IIa and IIb), 
that is, the estimation is independent from different experimental window for averaging. 
However, $r_{\rm tube}$ depends on which fit is adopted, i.e., it depends on fit-A or fit-B. 
We obtain $r_{\rm tube}= 0.2-0.3$ fm in the analysis by the fit-A, on the other hand,  we have $r_{\rm tube}= 0.4-0.7$ fm by the fit-B. 
Note that the fit-B faithfully reproduces the multiplicity dependence of the fluctuation measure observed by removing data with low multiplicity from fit.
Resulting $r_{\rm tube}$ is found to be very sensitive to the slope of the fluctuation measure $\sqrt{C_m}/\langle p_{\rm T}\rangle$. 
In addition to $r_{\rm tube}$, we compare $dn_{\rm ch}/dy$ dependence of $R_{\rm T}$ obtained and HBT radii observed by ALICE \cite{Aamodt:2010jj} in Fig.\ref{fig:tube_size}. 
Since parton-hadron duality is assumed,  $S_{\rm T}/\gamma$ is a scale variable. Therefore, $S_{\rm T}=\pi R_{\rm T}^2$ and $\gamma$ can not be independently determined. 
We solved this scaling by requiring that the transverse interaction area obtained by HBT observations be equal to $S_{\rm T}$ (where transverse expansion of the interaction region is ignored).
The resulting $\gamma$ is $2.20$, $2.42$, and $2.50$ for the case of type I, IIa and IIb, respectively. 
Although the absolute value of the size has been adjusted to HBT measurement, it is worth noting that the multiplicity dependence also reproduces the observation results. 
Furthermore, when the $dn_{\rm ch}/dy < 2\sim3$, calculations in the case of fit-B show that $r_{\rm tube} \approx R_{\rm T}$. 
This indicates that the number of tubes produced in that case is at most one.

We also show the multiplicity dependence of the $\lambda_g$ obtained in Fig.\ref{fig:lambda} with 3 data points of 
$\lambda_{\rm HBT}$ parameter in the HBT observation measured by ALICE Collaboration for charged hadrons. 
If our model correctly evaluates the correlation strength of the identical gluon generated from color flux tube, 
there will be a large gap between the correlation strength of gluon and that of the hadron in the final state as seen in fig.\ref{fig:lambda}. 
It is not clear at the present time how this gap is explained, but the hadronization processes may have some influence on the correlation strength. 
It may be meaningful to recall color reconnecting (CR) \cite{Sjostrand:2013cya} mechanism, 
which is successfully reproduce multiplicity dependence of  $\langle p_{\rm T} \rangle$ \cite{Liu:2016apq}. 
Although CR does not affect mean transverse momentum fluctuation itself \cite{Abelev:2014ckr}, it may be possibly contribute to the gap. 
\begin{table}
\begin{center}
\caption{
Values of parameter $\sigma$ and the exponent $n$ in eq.(\ref{eq:chaoticity}) 
obtained from the fit to ALICE data\cite{Abelev:2014ckr} 
on the event-by-event mean $\langle p_{\rm T}\rangle$ fluctuation.}
\label{tab:2}
\begin{indented}
\item[] \begin{tabular}{ c c c c c c c}
\br 
&\multicolumn{3}{l}{fit-A } &\multicolumn{3}{l}{fit-B } \\  \mr 
~type~~~ & ~~$\sigma$~~ & ~~$n$~~ & $\chi2$/dof & \qquad $\sigma$~~ & ~~$n$~~ & $\chi2$/dof\\ \mr 
I         & 0.755    & $-0.525$   &  16.6/27          & 0.096  & $-0.872$   &  3.27/25 \\ 
IIa       & 0.515    & $-0.555$   &  17.8/27          & 0.132 & $-0.788$    &  3.17/25 \\ 
IIb       & 0.535    & $-0.550$   &  17.7/27          & 0.150 & $-0.770$    &  3.08/25 \\ 
\br 
\end{tabular}
\end{indented}
\end{center} 
\end{table}

\section{Summary and concluding remarks}\label{sec:4}  
We have proposed a model of $p_{\rm T}$ fluctuation in multi-particle interaction processes and 
applied it to $p$-$p$ collisions at energy $\sqrt{s}=0.90$ TeV. Our model is characterized 
by the saturation momentum $Q_{\rm sat}$ which is typical transverse momentum for given energy 
and multiplicity class \cite{McLerran:2013una,Praszalowicz:2013fsa} in the geometrical scaling (GS) picture.  
If local parton-hadron duality provides good phenomenological description for multi-particle production, 
some signature of GS picture can be observed in the charged particle of the final state. 
This assumption is very strong request which connects the final and initial states of the multi-particle interaction process directly. 
A comment concerning the validity of the assumption of the parton-hadron duality may be in order here 
because some possible later-stage re-scattering effects such as resonance decays or jet fragmentations may potentially have influence on the final state.  
Note that the upper limit for the transverse momentum of the type IIa and IIb ($p_{\rm T}<$~10~GeV/c) is 
larger than that of the type I ($p_{\rm T}<$~4~GeV/c). Hence, some of the difference between type I and type II 
may be attribute to the contribution from jet fragments. 
However, there is no substantial difference between results of type I and II in the analysis of $\sqrt{C_m}/\langle p_{\rm T}\rangle$ except for the difference in the best-fit parameter values. 
This indicates that jet fragmentation may not contribute to $p_{\rm T}$ fluctuation measure at least for the acceptance $0.15 <p_{\rm T} < 2.0$ GeV/c. 
Resonance particles such as $\rho$ meson decaying into daughter particles may also induce correlations in the momentum space. 
Since various resonance particles each have many decay processes, it is very difficult to know exactly 
how much resonance decay contributes to the amount of correlation and fluctuation.
Hence, one needs to use some elaborate event simulations such as {\tt Therminator} \cite{Kisiel:2005hn} to investigate this issue. 
Although we cannot examine the influence in detail in this paper, 
it should be noted here that, a simulation with {\tt Therminator} revealed a negligible contribution of resonance 
decays to the $p_{\rm T}$ correlation/fluctuation \cite{Kisiel:2005hn}.  
This indicates that cancellation between contributions of various resonance particles can occur \cite{Broniowski:2005ae}. 
 
Under the local parton-hadron duality being established,  
an experimental fact that $p_{\rm T}$ spectrum of the final state hadron is reproduced by a 
Tsallis-type function means that the universal function of the GS is also given by the 
Tsallis-type function eq.(\ref{eq:universal_func}).  
Thus so-called slope parameter or temperature extracted from transverse spectra is related to the   
the saturation momentum $Q_{\rm sat}$. 
Since the single inclusive $p_{\rm T}$ distribution can be regarded as a convolution of the universal 
function ${\cal F}$ and 
the 
multiplicity distribution $P(n_{\rm ch})$, 
one must be careful in determining the 
two parameters $q$ and $\kappa$ appearing in 
eq.(\ref{eq:universal_func}).   
For $p$-$p$ collision at energy $\sqrt{s}=$0.90 TeV, we found $q=1.11\sim1.12$ and $\kappa=0.135\sim0.139$ (see table \ref{tab:1}),
while $Q_{\rm sat}$ varies with $dn_{\rm ch}/dy$ as shown in Fig.\ref{fig:QsRt}. 

Considering that there is a Bose-Einstein correlation (BEC) between two gluons generated from the same color flux tube,
we introduce a BEC-type correlation function into the two gluon spectra eq.(\ref{eq:BEC_correlation}).  
The correlation function includes two parameters which characterize the color flux tube as the gluon sources, 
i.e., the correlation strength $\lambda_g$ and the parameter $\sigma$ 
relating to the transverse size of the color flux tube $1/(\sqrt\sigma Q_{\rm sat})$. 
If such BEC correlations in the transverse momentum space survives until hadronization, 
the fluctuation measure of $p_{\rm T}$ such as $\sqrt{C_m}/\langle p_{\rm T} \rangle$ observed by ALICE Collaboration 
includes information about the size of the color flux tube\footnote{The Bose-Einstein correlations play a role as a source of fluctuation, but 
it can be also thought that the saturation momentum $Q_{\rm sat}$ itself fluctuates. 
If $Q_{\rm sat}$ fluctuates in event-by-event,  its variance is directly connected with $q$ parameter \cite{Wilk:2009nn} appearing in the universal function.}.  
By fitting the ALICE data on the event-by-event $\langle p_{\rm T} \rangle$ fluctuation with the theoretical curve obtained from our model,   
both transverse size of the color flux tube and the interaction area were evaluated as shown in Fig.\ref{fig:tube_size}. 
The transverse size of the color flux tube estimated is 0.4$\sim$ 0.7 fm (from fit-B) and the value depends on the event multiplicity. 
On the other hand, the transverse size of the interaction area $R_{\rm T}$ increases as the multiplicity increases, 
which is consistent with results of HBT measurements. 
It is interesting to investigate whether such properties found in this article 
can be obtained for different energy and system. 
We plan to address these issues by analyzing ALICE results on the 
$C_m/\langle p_{\rm T} \rangle$ for $p$+$p$ and Pb+Pb collisions at higher energies. 

\section*{Acknowledgement}
We would like to thank Kazunori Itakura 
for illuminating discussions on the fluctuation 
of the saturation momentum and 
on the origins of the fluctuation. 
\vspace*{10mm}

\section*{References}
\bibliographystyle{iopart-num} 
\bibliography{ptmultip17}

\end{document}